\documentclass[a4paper]{jpconf}
\usepackage{graphicx}
\usepackage[square,sort&compress,numbers]{natbib}
\begin{document}
\title{Gamma-Ray Astronomy from the Ground}

\author{Dieter Horns}

\address{University of Hamburg, Luruper Chaussee 159, 22761 Hamburg}

\ead{dieter.horns@physik.uni-hamburg.de}

\begin{abstract}
	 The observation of cosmic gamma-rays from the ground is based upon
	 the detection of gamma-ray initiated air showers. At energies  between
	 approximately $10^{11}$~eV and $10^{13}$~eV, the imaging air Cherenkov technique is a particularly
	 successful approach to observe gamma-ray sources with energy fluxes 
	 as low  as  $\approx 10^{-13}$ erg\,cm$^{-2}\,$s$^{-1}$. 
	 The observations of gamma-rays in this energy band probe particle
	 acceleration in astrophysical plasma conditions and are sensitive
	 to high energy phenomena beyond the standard
	 model of particle physics (e.g., self-annihilating or decaying dark matter,
	 violation of Lorentz invariance, mixing of photons with 
	 light pseudo-scalars). The current standing of the 
	 field and its major instruments 
	 are summarised briefly by presenting selected highlights. 
	 A new generation of ground based gamma-ray instruments is currently
	 under development. The perspectives and opportunities of these future
	 facilities will be discussed.
\end{abstract}

\section{Introduction}
 The field of ground based gamma-ray observations has been driven by the
 pioneering efforts to detect air Cherenkov light from extended air showers.
 The vivid history leading up to the discovery of the first cosmic gamma-ray
 source, the Crab nebula,  with the Whipple telescope and the subsequent
 evolution of mainly imaging air Cherenkov telescopes has been told by one of
 the pioneers himself in \cite{2012fura.book..143L}.  \\
 Currently, the ground based detection of gamma-rays is dominated by the
 imaging air Cherenkov technique which combines a collection area of $\approx
 10^5$~m$^2$ and a field of view of $\approx 5$~msrad with an event-by-event
 relative energy resolution better than 20~\% and an angular resolution better
 than $0.1^\circ$. Alternative approaches are explored to increase the size of the field of view
 (e.g., the water-Cherenkov detector HAWC) or increase the collection area to
 improve the sensitivity for ultra-high energy gamma-rays ($>100$ TeV) (see also section~\ref{section_future}).  \\
 With the very successful operation of the Fermi-LAT (in orbit since August
 2008, see also the review article in this volume), it has become feasible for the first time to
 study the energy spectra of gamma-ray sources with continuous coverage from
 100 MeV up to 100 TeV.\\
 In the following section, we summarise the status of the operating instruments
 and present selected highlights of the past two years (for prior results see
 reviews of previous TAUP conferences \cite{2012JPhCS.375e2020P,maier2013}).

\section{Status of currently operating major instruments}
 \subsection{High Altitude Water Cherenkov Observatory (HAWC)}
\begin{figure}
	 \begin{minipage}[t]{0.48\linewidth}
	 \includegraphics[width=\linewidth]{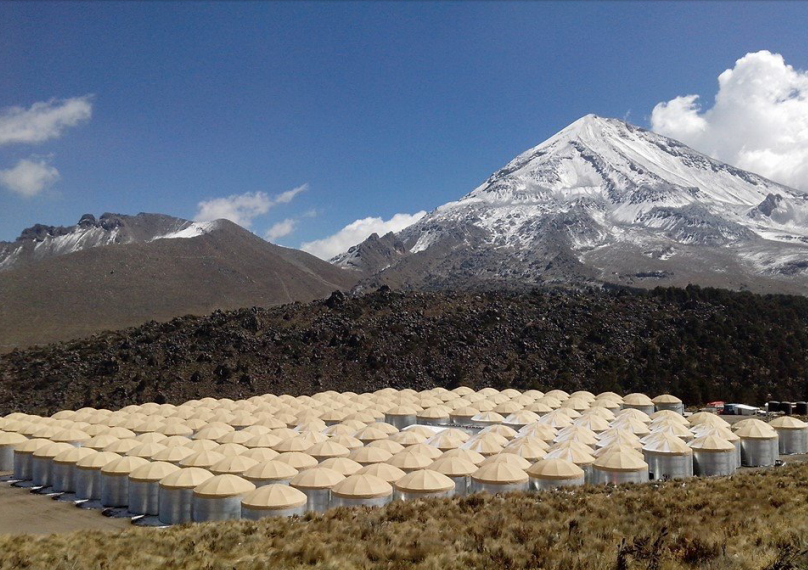}
 \caption{\label{HAWC} The High Altitude Water Cherenkov (HAWC) Observatory:
	 300 individual water Cherenkov detectors are installed over a surface
 area of 20~000 m$^2$ at an altitude of 4~100~m (adapted from \cite{2015arXiv150907851P}).}
\end{minipage}\hspace{2pc}%
\begin{minipage}[t]{0.48\linewidth}
		 \includegraphics[width=\linewidth]{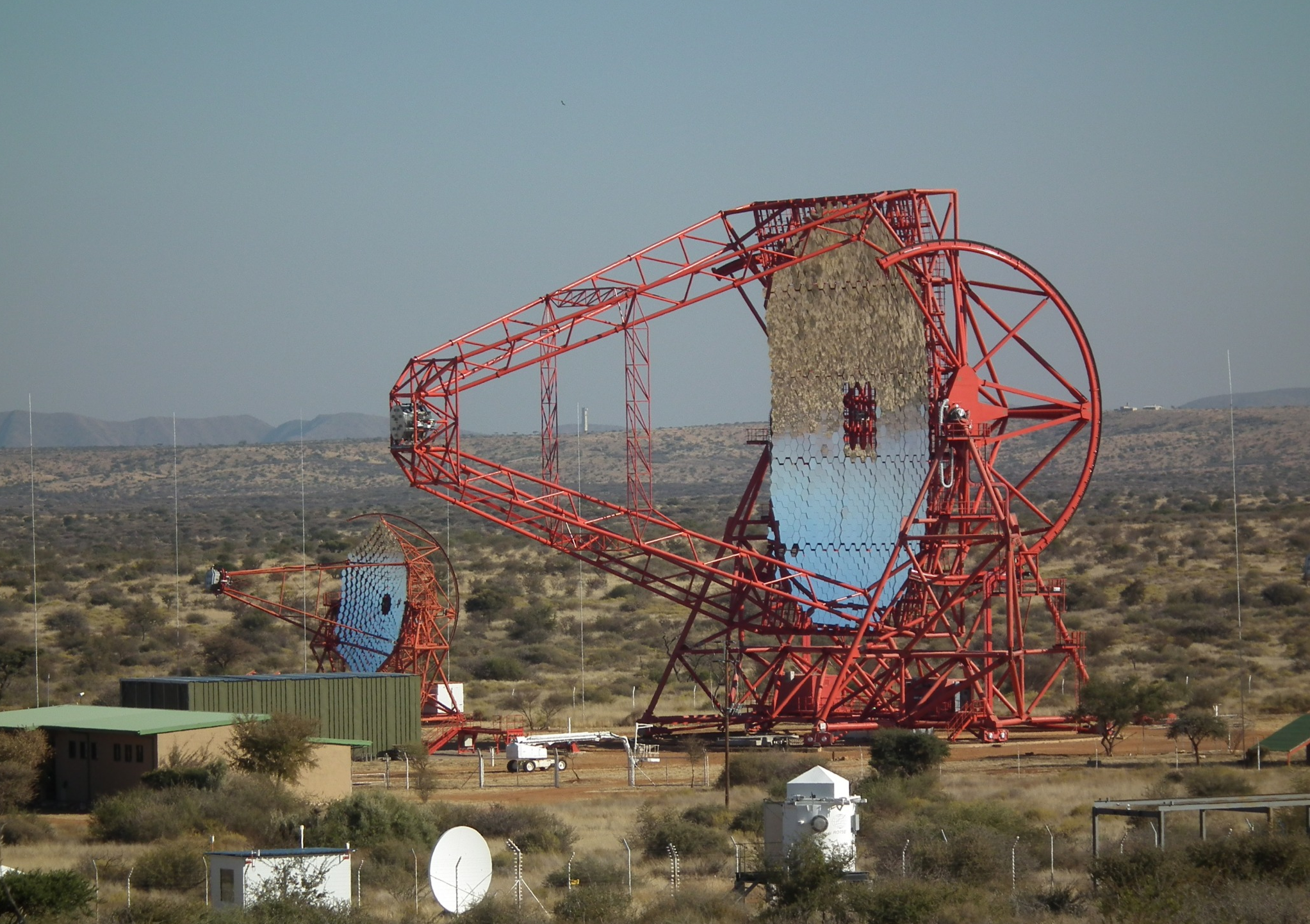}
		 \caption{\label{hessii} The recent addition to the H.E.S.S.
		 telescope array (four telescopes of the type seen on the
	 left): the fifth telescopes dwarfes the existing telescopes with a
 mirror surface of 600~m$^2$ (picture courtesy of the H.E.S.S. collaboration).}
\end{minipage} 
 \end{figure}

 The HAWC is designed to detect air showers via Cherenkov light generated by
 through-going charged particles in water tanks.  An array of 300 water tanks
 has been installed over the past three years (see Fig.~\ref{HAWC} for a
 picture of the installation) at an altitude of 4\,100~m a.s.l. in Mexico. At
 this high altitude, air showers can be detected that were initiated by
 gamma-rays with an energy above 100~GeV. The main array has been completed in
 2015 \cite{2015arXiv150907851P}. HAWC combines a low energy threshold,
 large collection area, $\approx 100$~\% duty cycle, and a large field of view
 of 5~srad. The sensitivity of the instrument after one year of operation is
 sufficient to detect any gamma-ray source above 2~TeV with an energy flux of
 $\approx 10^{-12}$~erg~cm$^{-2}$~s$^{-1}$ in the northern sky. This matches
 quite well the sensitivity of the air Cherenkov telescopes after a few ten
 hours of pointed observation time. The survey of the northern sky has already
 started with data taking during the commissioning phase \cite{2015arXiv150907851P,2015arXiv150905401A},
 demonstrating the opportunities of a large field of view instrument,
 i.e., an unbiased survey and the sensitivity to detect extended emission
 difficult to achieve with narrow field of view instruments.\\
 Another important ingredient of the scientific shopping list are transient
 events in the gamma-ray sky, e.g., gamma-ray bursts which should be detectable
 with HAWC.  
 \subsection{High Energy Stereoscopic System (H.E.S.S.): Phase II}
 The H.E.S.S. array in its current configuration (\textit{Phase II}), consists
 of four telescopes (100 m$^2$ mirror surface area) located at the corners of a
 square with 120~m side length and a considerably larger telescope of $\approx
 600$~m$^2$ surface area in the center (see Fig.~\ref{hessii} for a 
 picture of the installation). The large telescope (\textit{CT~5}) was
 inaugurated in 2012 and has considerably lowered the detection threshold below
 100~GeV and has improved the reconstruction accuracy of so-called hybrid
 events. These events are triggered by the large telescope in conjunction with at least one smaller
 telescope. The small telescopes are currently
 undergoing an extensive update of the camera readout to improve the
 performance at low energies \cite{2015arXiv150901232G}. 
 \subsection{MAGIC$^2$: two 17~m-telescopes}
\begin{figure}[h]
	\begin{minipage}[t]{0.48\linewidth}
\includegraphics[width=\linewidth]{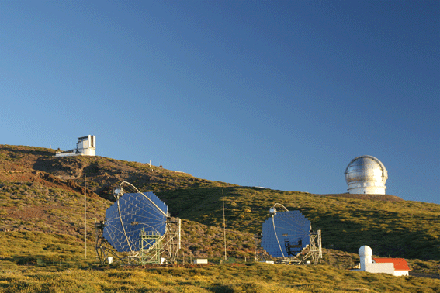}
\caption{\label{2magics} The two MAGIC telescopes located on the peak of the Roque de los Muchachos on La Palma (2200 m asl), picture courtesy of the MAGIC collaboration.}
\end{minipage}\hspace{2pc}%
\begin{minipage}[t]{0.48\linewidth}
\includegraphics[width=\linewidth]{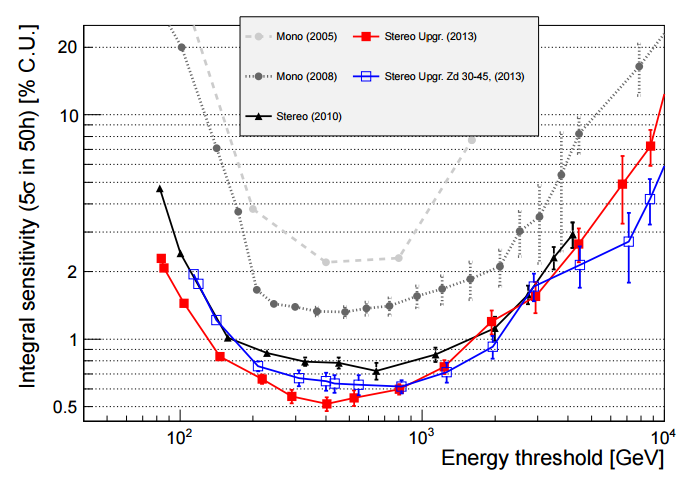}
\caption{\label{magic2_perf} Evolution of the integrated flux sensitivity of the MAGIC telescopes 
	following upgrades: 
	the most recent update of the camera and trigger
system has lead to the improvement from the curve with the black, filled
triangles to the red, filled squares. Figure from \cite{2016APh....72...76A}.}
\end{minipage} 
\end{figure}

 The MAGIC collaboration has successfully constructed imaging air Cherenkov
 telescopes featuring a number of innovative technologies including a partial
 carbon fibre frame to reduce the weight and therefore increase the achievable
 re-pointing speed, new composite mirror technology, as well as hemispherical
 photo-multiplier tubes to improve the optical detection efficiency. The most
 recent updates have been the addition of a second telescope (see
 Fig.~\ref{2magics}) and the update of the cameras to achieve a more uniform
 performance for the stereoscopic observation of air showers \cite{2016APh....72...61A}. The improvement
 in performance of the telescope and its resulting sensitivity has been
 spectacular, pushing 
 the initial flux sensitivity above 100 GeV down by as much as one order of magnitude. The
 recent upgrade of the camera and trigger system has improved the sensitivity
 by a factor of two at the energy threshold of 70~GeV (see
 Fig.~\ref{magic2_perf}) \cite{2016APh....72...76A}. \\ 
 The MAGIC collaboration continues to pioneer new techniques to improve the
 performance of the telescopes. Recently, they demonstrated successfully that
 LIDAR backscattering results can be used to correct the air shower
 observations taken under conditions \cite{2014arXiv1403.3591F} which are usually rejected because of poor
 transparency of the atmosphere.  This is an important step to both improve the
 energy reconstruction which suffer systematic uncertainties because of
 variations in the atmosphere and to recover observation time. The new technique is of relevance
 for observations taking place during unique flux states of variable sources or
 during multi-wavelength campaigns where the weather conditions were not cooperative.

 \subsection{VERITAS: four 12~m-telescopes}
  The VERITAS array has been operational since 2012 with new photo-multiplier
  tubes with an increased detection efficiency \cite{2013arXiv1308.4849D}. The energy threshold
  for triggering on gamma-ray induced air showers 
  has been reduced 30~\%. At the same time, the
  sensitivity has been improved because of larger number of photo electrons
  that can be used to characterise the image and to separate cosmic-ray induced
  background. \\
  Recently, the VERITAS array started operation  during (partial) moon
  light in order to increase the observation time available. Two different
  modes of observation have been successfully
  tested \cite{2015arXiv150807186G}: reducing the high voltage and adding UV bandpass filters in front of
  the PMTs. The resulting performance deteriorates only slightly with reduced
  high voltage while it permits for observations with a moon illumination up to
  65~\%. For observations with even more moonlight, UV filters can reduce the
  light load on the PMTs, however at the expense of an increased energy
  threshold and reduced sensitivity (roughly a factor of two with respect to
  nominal operation). These results are consistent with the 
  previous studies carried out with the MAGIC telescope where routine moonlight
  observations are scheduled \cite{2007astro.ph..2475M}. 

  \section{Scientific highlights  since 2013}
  \begin{figure}
	   \includegraphics[width=0.95\linewidth]{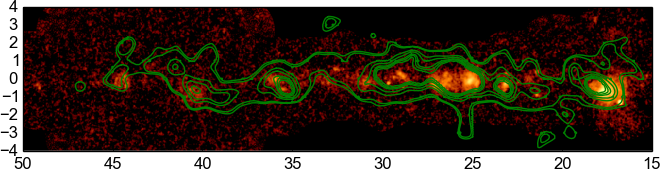}
	   \caption{\label{hawc_hess} 
Color-scale: section of the first Galactic quadrant from the H.E.S.S. survey (adapted from
\cite{deilicrc}); overlay: preliminary
	   contours tracing iso-significance from HAWC observations taken during commissioning 
(adapted from \cite{2015arXiv150907851P}).}
  \end{figure}
  \subsection{Surveying the Galactic plane}
    The population of charged cosmic rays arriving at Earth is commonly
    considered to be the result of localised acceleration processes  -- 
    mainly at shock fronts formed by supernova shells, colliding
    winds etc.   and in or near to magneto-spheres of neutron stars. \\
    The Galactic plane is therefore an ideal target for a survey to take
    inventory of nearby and powerful cosmic-ray accelerators.  The H.E.S.S.
    telescope array has been used to carry out an extensive survey of the inner
    part of the Galactic plane  for a total
    of 3~000 hours \cite{deilicrc}. The initial shallow survey carried out during the first years
    of operation resulted in the landmark 
    discovery of a new population of gamma-ray sources \cite{2006ApJ...636..777A} with many of the objects
    lacking obvious counterparts at other wavelengths. The current survey has improved both in coverage
    (longitude range: $l=-110^\circ \ldots +70^\circ$ in comparison to $l=-30^\circ\ldots +30^\circ$) 
    and in depth, reaching a minimum detectable flux of $<1~\%$ in the inner Galaxy initially surveyed (previously
    $2~\%$).   \\
    The first example of a gamma-ray emitting source which belongs to the most abundant type in the
    Galaxy type  had already been
    discovered before the Galactic plane survey with H.E.S.S. started. During observations from 1999-2000, 
    the first unidentified TeV gamma-ray source (TeV J2032+4130)  had been
    discovered with the HEGRA Cherenkov telescopes \cite{2002A&A...393L..37A}.  The discovery was
    later confirmed with MAGIC \cite{2008ApJ...675L..25A} and VERITAS observations \cite{2014ApJ...783...16A}.  \\
    The majority of unidentified gamma-ray objects later discovered in the Galactic plane
    surveys share similar properties with the prototypical TeV~J2032+4130: 
    both, spatial extension as well as luminosity
    at gamma-rays are typically larger than  potential counterpart candidates at X-rays.\\
     The common interpretation of these objects as evolved pulsar wind
    nebula systems  assumes that the pulsar releases over its life time an 
    expanding magnetised relativistic plasma. The decline of the magnetic field over time 
    favors inverse Compton gamma-ray emission
    over synchrotron emission which explains the lack of obvious counterparts. This interpretation 
    has been supported in individual cases 
    by the discovery of pulsed emission with the Fermi LAT instrument as 
    well as with the detection of faint X-ray nebula emission around known radio pulsars.\\
    In this scenario, the X-ray emitting nebula and the gamma-ray nebula do not
    necessarily have a similar morphology because of the different energy of
    the underlying electron population responsible for the emission in the
    respective energy band.  A population study of the gamma-ray pulsar wind nebulae  has
    been presented recently \cite{klepsericrc}. 
    This sample of similar objects indicates that the efficiency (ratio
    of gamma-ray luminosity to spin-down power of the pulsar) increases with increasing age of the  system:
    the gamma-ray luminosity is a proxy of the integrated spin-down power of the system while
    X-rays are emitted by the recently injected electrons. \\
    %
%    The Galactic plane survey with the H.E.S.S. telescopes includes a total of $\approx 3000$~hrs of observation
%    and results in the detection of 64 sources and 13 complex source regions. The release of a first catalogue of Galactic gamma-ray
%    sources has been announced for the past years. The delay  of its release indicates the complexity of the task. \\
    %
    The analysis of 
    Galactic plane data taken with HAWC during its commissioning phase \cite{2015arXiv150905401A} 
    demonstrates the complementarity of 
    a wide field instrument to narrow field imaging telescopes. The large field of view simplifies the analysis of 
    extended emission regions. Combining the data in overlapping energy ranges is an exciting opportunity
    to disentangle extended emission from more point-like contributions (see Fig.~\ref{hawc_hess} for
    an overlay of HAWC significance iso-contours with the H.E.S.S. data in an overlapping region in the 
    first quadrant of the Galactic plane). \\
    Even though the detection of source emission extending beyond the field of
    view of Cherenkov telescopes is challenging, a recent result put forward by
    the H.E.S.S. collaboration claims the  detection of  diffuse
    emission along the Galactic plane \cite{2014PhRvD..90l2007A}.
 After masking sources detected in the inner Galaxy ($|l|<70^\circ$), a diffuse emission centered on the Galactic
    equator with a full width half maximum of $\approx 0.75^\circ$ is detected. 
The observed intensity is a factor 2-3 larger than
    expected in a simple model extrapolating Galactic emission models into the regime above 100~GeV. 
Future studies in junction with 
    data taken with the Fermi-LAT and HAWC will help to  clarify the origin of the diffuse emission.

    \subsection{Gamma-ray emission from the Galactic center}
     The Galactic center and the inner 100~pc are an exceptional region for the study of non-thermal emission and 
     diagnostics of the relativistic plasma. A steady gamma-ray source 
     consistent with the position of Sgr~A$^*$ was discovered with ground-based observations 
\cite{2004ApJ...606L.115T,2004A&A...425L..13A,2004ApJ...608L..97K}.
     Subsequent deeper observations 
     with the H.E.S.S. telescopes revealed extended gamma-ray emission along the Galactic ridge, tracing the distribution
     of molecular gas \cite{2006Natur.439..695A}. 
The most recent results of even deeper observations resolve in more detail the distribution of
     cosmic-rays in the inner 200~pc \cite{aviana_icrc}: The energy density of cosmic rays 
     in the inner 200~pc is approximately an order of magnitude larger than the average local 
cosmic-ray energy density. 
     This density falls off $\propto 1/r$ and the energy spectrum follows a power-law that 
extends without an indication for a cut-off well beyond 10~TeV.  This in turn
implies the acceleration and release of particles well beyond 100 TeV from the
Galactic center.\\
     Deeper observations have also revealed the presence of structures in the
morphology which follow the well-known radio feature known as the arc \cite{lemiere_icrc,smith_icrc}.
 Future
studies and interpretational work will provide a better insight in the particle
acceleration and
     cosmic ray release in the inner Galaxy.
    
    \subsection{Dark matter searches with gamma-rays}
    The indirect search for gamma-rays from self-annihilating Dark matter
continues with a focus on observations of dwarf spheroidals and
from the vicinity of the Galactic center. The sensitivity continues
to improve  with longer exposures, new instrumentation, and through
combination of different observations.

\begin{figure}
\begin{minipage}[t]{0.52\linewidth}
 \includegraphics[width=\linewidth]{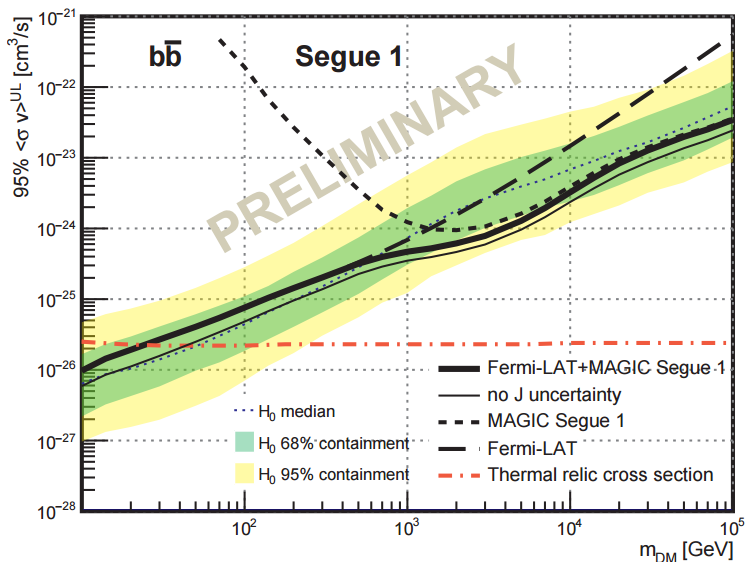}
 \caption{Combined constraint (confidence level 95~\%) 
on the cross section for self-annihilating
dark matter from the dwarf spheroidal galaxy Segue 1 for a $b\bar b$
final state. The $H_0$ bands indicate containment of 
limits from simulated/blank field data for comparison.
 The combined limit using both Fermi-LAT and MAGIC data 
improves the sensitivity in the overlapping mass range (800~GeV-8 TeV)
and improves the mass range covered, adapted from \cite{2015arXiv150805827R}.
\label{magic_segue}}
\end{minipage}
\begin{minipage}[t]{0.46\linewidth}
 \includegraphics[width=\linewidth]{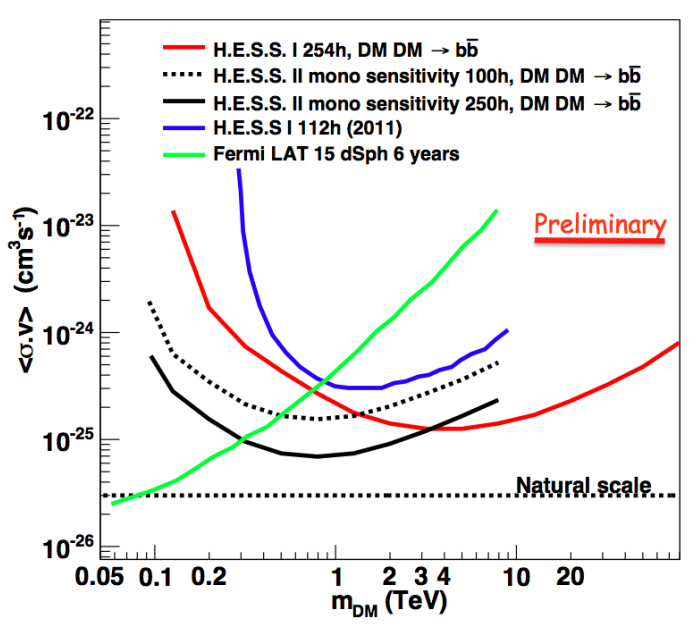}
\begin{center}
\begin{minipage}{0.9\linewidth}
 \caption{Expected constraint on the dark matter self-annihilation cross section
from observations of the region close to the Galactic center 
with the H.E.S.S. phase II telescope (black solid and dashed line). For comparison, existing constraints from Galactic center observations (HESS I, blue
line) and 
a new result (red line) based upon 254~hours of observation are shown
 \cite{lefranc_icrc}.
\label{hess_gc}}
\end{minipage}
\end{center}
\end{minipage}
\end{figure}

 The combination of observational data from Fermi-LAT with MAGIC
on dwarf galaxies has improved the energy coverage and sensitivity in comparison
to the individual analysis \cite{2015arXiv150805827R}(Fig.~\ref{magic_segue}). 
Deeper observations of the Galactic center halo with H.E.S.S. 
have improved the sensitivity in comparison to previous observations. With upcoming data using the large central
telescope, the sensitivity will mainly improve for the WIMP mass range
below a few TeV \cite{lefranc_icrc} (see Fig.~\ref{hess_gc}). 

  \subsection{Gamma-rays from active galactic nuclei and gamma-ray propagation}
 The catalogue of gamma-ray emitting active galactic nuclei (AGN) continues to
grow and to include different types of AGN.  The majority of the sources belong
however to the so-called high energy peaked BL Lac type, sources without strong line
emission features in the optical spectra and a broad band spectral energy
distribution with a maximum in the X-ray and one at gamma-ray energies.\\
Recently a number of flat-spectrum radio quasars have been observed to emit
gamma-rays and the red-shift of some  these recently with MAGIC discovered gamma-ray emitting objects 
(PKS~1441+25 \cite{2015ATel.7416....1M,2015arXiv151204435M,2015ApJ...815L..22A} 
and B0218+357 \cite{2014ATel.6349....1M,2015arXiv150804580S}) is close to $z\approx 1$.\\
A striking property of the high energy emission from AGN is the dramatic
variability observed at the highest energies.  The short time-scales of
variability impose severe constraints on the spatial extension of the emitting
volume. The recent MAGIC observation
of flares from the head-tail radio galaxy IC~310 with a variability time scale of approximately 4~minutes, 
is particularly puzzling as it requires an emission region more compact (by a factor of $\approx 5$) 
than the Schwarzschild radius of the 
super massive black hole \cite{2015arXiv150805031E}. \\
Another independent inference of a potentially even more compact size of the emission region in Blazars
is possible using the 
effect of gravitational micro lensing \cite{2015arXiv150701092V}. 
The observation of the time-delays and the ratios of the flux amplifications 
provides insights into the velocity of emitting volume as well as on its size. Recently, the observation
of B0218+357 with MAGIC followed up on a flare detected with Fermi-LAT and lead to the discovery of the 
signal delayed by gravitational lensing \cite{2015arXiv150804580S}.  \\
 The large red shift of gamma-ray sources and the observation of gamma-rays up
to and beyond 20 TeV from AGN in general is surprising given that the gamma-ray
flux from large distance sources is expected to be attenuated because of the
effect of pair-production. Possible explanations for a reduced attenuation have
been put forward and include the possibility of AGN to be powerful accelerators
of ultra-high energy cosmic-rays with gamma-ray production through
inter-galactic cascading  as well as more exotic
proposals 
like conversion of photons with low-mass axion-like particles \cite{2007PhRvD..76l1301D}, 
or violation of Lorentz invariance \cite{1999ApJ...518L..21K},
for an overview of gamma-ray propagation see e.g., \cite{HornsJacholkowska}.
\\

  \section{Towards the future}
\label{section_future}
The future of ground based gamma-ray observations is starting with the planning
and construction of the next generation of instruments.  The Cherenkov
telescope array (CTA) project \cite{2013APh....43....3A} is close to finalising the design of two future
arrays of Cherenkov telescopes installed on the northern and southern
hemisphere. While the site selection continues, the technical design of the
three telescope types (small: mirror area 12~$\mathrm{m}^2$, field of view
9$^\circ$, medium: 110~$~\mathrm{m}^2$, 7$^\circ$ and large:
450~$\mathrm{m}^2$, 4.5$^\circ$) progresses quickly as it builds upon the experience with the current
generation of telescopes. The aim of the project is a general improvement of the sensitivity by one order of magnitude and 
an extension of the energy reach, including both the low energies ($E<100$~GeV) and high energies ($E>10$~TeV). \\
In order to achieve an even  lower energy threshold, 
the MACE telescope (area of 340~m$^2$, 4$^\circ$) is under construction 
at 4270~m altitude and is planned to start regular observations in 2017 with an energy threshold of 20~GeV
\cite{2013ASInC...9..100Y}.
\\
At higher energies and going beyond 100~TeV, the current imaging concept is expensive to instrument sufficiently large 
areas to achieve sensitivity to detect the expected low fluxes. The LHAASO project will combine 
non-imaging and imaging techniques at a high altitude to achieve a broad energy coverage and reach high energies with a 
collection area of 1~km$^2$ \cite{2010cosp...38.2322C}.\\
 The TAIGA project is using non-imaging air Cherenkov detectors employed on a large surface
(up to 10~km$^2$) in combination with imaging telescopes \cite{2014APh....56...42T}.
 Both experiments are in the build-up phase and will start taking
data with sufficient sensitivity to detect first sources in the next years. 
\bibliography{taup}

\providecommand{\newblock}{}
\begin{thebibliography}{10}
\expandafter\ifx\csname url\endcsname\relax
  \def\url#1{{\tt #1}}\fi
\expandafter\ifx\csname urlprefix\endcsname\relax\def\urlprefix{URL }\fi
\providecommand{\eprint}[2][]{\url{#2}}

\bibitem{2012fura.book..143L}
{Lorenz} E and {Wagner} R 2012 {\em {Very-High Energy Gamma-Ray Astronomy: A
  23-Year Success Story in Astroparticle Physics}\/} p 143

\bibitem{2012JPhCS.375e2020P}
{Paneque} D 2012 {\em Journal of Physics Conference Series\/} {\bf 375} 052020

\bibitem{maier2013}
{Maier} G 2014 {\em Physics of the Dark Universe\/} {\bf 4} 1

\bibitem{2015arXiv150907851P}
{Pretz} J {\em et~al.\/} 2015 {\em ArXiv e-prints\/} (\textit{Preprint}
  \eprint{1509.07851})

\bibitem{2015arXiv150905401A}
{Abeysekara} A~U, {Alfaro} R, {Alvarez} C {\em et~al.\/} 2015 {\em ArXiv
  e-prints\/} (\textit{Preprint} \eprint{1509.05401})

\bibitem{2015arXiv150901232G}
{Giavitto} G, {Ashton} T, {Balzer} A {\em et~al.\/} 2015 {\em ArXiv e-prints\/}
  (\textit{Preprint} \eprint{1509.01232})

\bibitem{2016APh....72...61A}
{Aleksi{\'c}} J, {Ansoldi} S, {Antonelli} L~A {\em et~al.\/} 2016 {\em
  Astroparticle Physics\/} {\bf 72} 61--75 (\textit{Preprint}
  \eprint{1409.6073})

\bibitem{2016APh....72...76A}
{Aleksi{\'c}} J, {Ansoldi} S, {Antonelli} L~A {\em et~al.\/} 2016 {\em
  Astroparticle Physics\/} {\bf 72} 76--94 (\textit{Preprint}
  \eprint{1409.5594})

\bibitem{2014arXiv1403.3591F}
{Fruck} C, {Gaug} M, {Zanin} R, {Dorner} D, {Garrido} D, {Mirzoyan} R, {Font} L
  and {for the MAGIC Collaboration} 2014 {\em ArXiv e-prints\/}
  (\textit{Preprint} \eprint{1403.3591})

\bibitem{2013arXiv1308.4849D}
{D~B~Kieda for the VERITAS Collaboration} 2013 {\em ArXiv e-prints\/}
  (\textit{Preprint} \eprint{1308.4849})

\bibitem{2015arXiv150807186G}
{Griffin} S and {for the VERITAS Collaboration} 2015 {\em ArXiv e-prints\/}
  (\textit{Preprint} \eprint{1508.07186})

\bibitem{2007astro.ph..2475M}
{MAGIC Collaboration} and {Albert} J 2007 {\em ArXiv Astrophysics e-prints\/}
  (\textit{Preprint} \eprint{astro-ph/0702475})

\bibitem{deilicrc}
{Deil} C, {Brun} F, {Carrigan} S {\em et~al.\/} 2015 {\em The 34th
  international cosmic ray conference\/} Proc. of Science

\bibitem{2006ApJ...636..777A}
{Aharonian} F, {Akhperjanian} A~G, {Bazer-Bachi} A~R {\em et~al.\/} 2006 {\em
  ApJ\/} {\bf 636} 777--797 (\textit{Preprint} \eprint{astro-ph/0510397})

\bibitem{2002A&A...393L..37A}
{Aharonian} F, {Akhperjanian} A, {Beilicke} M {\em et~al.\/} 2002 {\em A\&A\/}
  {\bf 393} L37--L40 (\textit{Preprint} \eprint{astro-ph/0207528})

\bibitem{2008ApJ...675L..25A}
{Albert} J, {Aliu} E, {Anderhub} H {\em et~al.\/} 2008 {\em ApJL\/} {\bf 675}
  L25--L28 (\textit{Preprint} \eprint{0801.2391})

\bibitem{2014ApJ...783...16A}
{Aliu} E, {Aune} T, {Behera} B {\em et~al.\/} 2014 {\em ApJ\/} {\bf 783} 16
  (\textit{Preprint} \eprint{1401.2828})

\bibitem{klepsericrc}
{Klepser} S {\em et~al.\/} 2015 {\em The 34th international cosmic ray
  conference\/} Proc. of Science

\bibitem{2014PhRvD..90l2007A}
{Abramowski} A, {Aharonian} F, {Ait Benkhali} F {\em et~al.\/} 2014 {\em Phys.
  Rev. D\/} {\bf 90} 122007 (\textit{Preprint} \eprint{1411.7568})

\bibitem{2004ApJ...606L.115T}
{Tsuchiya} K, {Enomoto} R, {Ksenofontov} L~T {\em et~al.\/} 2004 {\em ApJL\/}
  {\bf 606} L115--L118 (\textit{Preprint} \eprint{astro-ph/0403592})

\bibitem{2004A&A...425L..13A}
{Aharonian} F, {Akhperjanian} A~G, {Aye} K~M {\em et~al.\/} 2004 {\em A\&A\/}
  {\bf 425} L13--L17 (\textit{Preprint} \eprint{astro-ph/0406658})

\bibitem{2004ApJ...608L..97K}
{Kosack} K, {Badran} H~M, {Bond} I~H {\em et~al.\/} 2004 {\em ApJL\/} {\bf 608}
  L97--L100 (\textit{Preprint} \eprint{astro-ph/0403422})

\bibitem{2006Natur.439..695A}
{Aharonian} F, {Akhperjanian} A~G, {Bazer-Bachi} A~R {\em et~al.\/} 2006 {\em
  Nature\/} {\bf 439} 695--698 (\textit{Preprint} \eprint{astro-ph/0603021})

\bibitem{aviana_icrc}
{Viana} A {\em et~al.\/} 2015 {\em The 34th international cosmic ray
  conference\/} Proc. of Science

\bibitem{lemiere_icrc}
{Lemiere} A {\em et~al.\/} 2015 {\em The 34th international cosmic ray
  conference\/} Proc. of Science

\bibitem{smith_icrc}
{Smith} A {\em et~al.\/} 2015 {\em The 34th international cosmic ray
  conference\/} Proc. of Science

\bibitem{2015arXiv150805827R}
{Rico} J, {Wood} M, {Drlica-Wagner} A, {Aleksi{\'c}} J, {for the MAGIC
  Collaboration} and {the Fermi-LAT Collaboration} 2015 {\em ArXiv e-prints\/}
  (\textit{Preprint} \eprint{1508.05827})

\bibitem{lefranc_icrc}
{Lefranc} V {\em et~al.\/} 2015 {\em The 34th international cosmic ray
  conference\/} Proc. of Science

\bibitem{2015ATel.7416....1M}
{Mirzoyan} R 2015 {\em The Astronomer's Telegram\/} {\bf 7416}

\bibitem{2015arXiv151204435M}
{MAGIC Collaboration} {\em et~al.\/} 2015 {\em ArXiv e-prints\/}
  (\textit{Preprint} \eprint{1512.04435})

\bibitem{2015ApJ...815L..22A}
{Abeysekara} A~U, {Archambault} S, {Archer} A {\em et~al.\/} 2015 {\em ApJL\/}
  {\bf 815} L22 (\textit{Preprint} \eprint{1512.04434})

\bibitem{2014ATel.6349....1M}
{Mirzoyan} R 2014 {\em The Astronomer's Telegram\/} {\bf 6349}

\bibitem{2015arXiv150804580S}
{Sitarek} J, {Becerra Gonz{\'a}lez} J, {Dominis Prester} D {\em et~al.\/} 2015
  {\em ArXiv e-prints\/} (\textit{Preprint} \eprint{1508.04580})

\bibitem{2015arXiv150805031E}
{Eisenacher Glawion} D, {Sitarek} J, {Mannheim} K, {Colin} P, {for the MAGIC
  Collaboration} {\em et~al.\/} 2015 {\em ArXiv e-prints\/} (\textit{Preprint}
  \eprint{1508.05031})

\bibitem{2015arXiv150701092V}
{Vovk} I and {Neronov} A 2015 {\em ArXiv e-prints\/} (\textit{Preprint}
  \eprint{1507.01092})

\bibitem{2007PhRvD..76l1301D}
{de Angelis} A, {Roncadelli} M and {Mansutti} O 2007 {\em Phys. Rev. D\/} {\bf
  76} 121301 (\textit{Preprint} \eprint{0707.4312})

\bibitem{1999ApJ...518L..21K}
{Kifune} T 1999 {\em ApJL\/} {\bf 518} L21--L24 (\textit{Preprint}
  \eprint{astro-ph/9904164})

\bibitem{HornsJacholkowska}
{Horns} D and {Jacholkowska} A 2016 {\em Comptes Rendus Physique\/} {\bf 17}

\bibitem{2013APh....43....3A}
{Acharya} B~S, {Actis} M, {Aghajani} T {\em et~al.\/} 2013 {\em Astroparticle
  Physics\/} {\bf 43} 3--18

\bibitem{2013ASInC...9..100Y}
{Yadav} K, {Yadav} K~K, {Bhatt} N, {Chouhan} N, {Sikder} S~S, {Behere} A,
  {Pithawa} C~K, {Tickoo} A~K, {Rannot} R~C, {Bhattacharyya} S, {Mitra} A~K and
  {Koul} R 2013 {\em Astronomical Society of India Conference Series\/} ({\em
  Astronomical Society of India Conference Series\/} vol~9)

\bibitem{2010cosp...38.2322C}
{Cao} Z, {Bi} X~J, {Cao} Z {\em et~al.\/} 2010 {\em 38th COSPAR Scientific
  Assembly\/} ({\em COSPAR Meeting\/} vol~38) p~2

\bibitem{2014APh....56...42T}
{Tluczykont} M, {Hampf} D, {Horns} D, {Spitschan} D, {Kuzmichev} L, {Prosin} V,
  {Spiering} C and {Wischnewski} R 2014 {\em Astroparticle Physics\/} {\bf 56}
  42--53 (\textit{Preprint} \eprint{1403.5688})

\end{thebibliography}
 \end{document}